\documentclass[11pt]{article}
\usepackage{balance}  % for  \balance command ON LAST PAGE  (only there!)
\usepackage{times}
\usepackage{fullpage}
\usepackage{graphicx}
\usepackage{color}
\usepackage{subfigure}
\usepackage{multirow}
\usepackage{amsfonts}
\usepackage{multicol}
\usepackage{authblk}
\usepackage{amsthm}
\usepackage{color}
\usepackage{soul}

\usepackage{wrapfig}
\usepackage[noend]{algorithm}
\usepackage{algorithmic}
\usepackage{distribalgo}

\definecolor{light-gray}{gray}{0.95}
\sethlcolor{light-gray}

\newcommand{\Config}{\ensuremath{\textit{Config}}}
\newcommand{\Command}{\ensuremath{\textit{Command}}}
\newcommand{\branches}{\ensuremath{\textit{branches}}}
\newcommand{\curConf}{\ensuremath{\textit{curConf}}}
\newcommand{\next}{\ensuremath{\textit{next}}}
\newcommand{\trunk}{\ensuremath{\textit{trunk}}}

\newcommand{\Nat}{\ensuremath{\mathbb{N}^{\ge 0}}}

\newcommand{\seqOf}{\ensuremath{\textit{seqOf}}}

\newcommand{\members}{\ensuremath{\textit{members}}}
\newcommand{\view}{\ensuremath{\textit{view}}}
\newcommand{\status}{\ensuremath{\textit{status}}}
\newcommand{\learnNext}{\ensuremath{\textit{learnNext}}}
\newcommand{\joinConf}{\ensuremath{\textit{joinConf}}}

\newcommand{\propose}{\ensuremath{\textsl{propose}}}
\newcommand{\learn}{\ensuremath{\textsl{learn}}}
\newcommand{\join}{\ensuremath{\textsl{join}}}

\newcommand{\recon}{\ensuremath{\textsl{recon}}}
\newcommand{\newconf}{\ensuremath{\textsl{new-conf}}}
\newcommand{\ready}{\ensuremath{\textsl{ready}}}

\newcommand{\rrsm}{\ensuremath{\textsc{r-rsm}}}
\newcommand{\nrrsm}{\ensuremath{\textsc{nr-rsm}}}

\newenvironment{item*}%
 {\begin{itemize}%
   \setlength{\itemsep}{-5pt}%
   \setlength{\parsep}{-5pt}%
   \setlength{\topsep}{-5pt}}%
 {\end{itemize}}

\begin{document}

%\title{FRAPP\'E: Fast Replication Platform for Elastic Services}
%\title{Speculative Reconfiguration for Elastic State Machine Replication}

\title{Reconfigurable State Machine Replication from Non-Reconfigurable Building Blocks}

% \title{Reconfigurable State Machine Replication from\\ Non-Reconfigurable Building Blocks\\
% {\normalsize \textbf{Algorithms, Applications, and Evaluation}}}

%\numberofauthors{6}
\author[1]{Vita Bortnikov}
\author[1]{Gregory Chockler}
\author[2]{Dmitri Perelman}
\author[1]{Alexey Roytman}
\author[1]{Shlomit Shachor}
\author[1]{Ilya Shnayderman}

\affil[1]{IBM Research} 
\affil[2]{Technion, Israel Institute of Technology}

\date{}

\maketitle
\thispagestyle{empty}

\newtheorem{thm}{Theorem}
\newtheorem{lemma}{Lemma}
\newtheorem{property}{Property}
\newtheorem{assumption}{Assumption}[section]
\newtheorem{observation}{Observation}[section]
\newtheorem{invariant}{Invariant}[section]
\newtheorem{definition}{Definition}[section]

% -----------------------------
% Ilya's cut and paste
\newcommand{\tup}[1]{%
        \relax\ifmmode
	           \langle #1 \rangle%
        \else
                $\langle$#1$\rangle$%
        \fi
}

\newcommand{\cnfgEst}{\ensuremath{\textit{cnfgEst}}}
\newcommand{\ballotNum}{\ensuremath{\textit{ballotNum}}}
\newcommand{\acceptedVal}{\ensuremath{\textit{acceptedVal}}}
\newcommand{\decidedVal}{\ensuremath{\textit{decidedVal}}}
\newcommand{\Proposal}{\ensuremath{\textit{Proposal}}}
\newcommand{\AcceptedValue}{\ensuremath{\textit{AcceptedValue}}}
\newcommand{\DecisionCandidate}{\ensuremath{\textit{DecisionCandidate}}}
\newcommand{\DecisionCandidates}{\ensuremath{\textit{DecisionCandidates}}}
\newcommand{\Set}{\ensuremath{\textit{Set}}}
 
\newfloat{interface}{htbp}{interface}
\floatname{interface}{Interface}

\newfloat{algo}{htbp}{algo}
\floatname{algo}{Algorithm}

\definecolor{Green}{rgb}{0.1,0.4,0.1}
\definecolor{Blue}{rgb}{0.1,0.1,0.6}
\definecolor{DarkBlue}{rgb}{0.2,0.2,0.5}
\definecolor{Cyan}{rgb}{0.5,0.2,0.5}
\newcommand{\comment}[1]{{\color{Green}{$\rhd$ #1}}}
\newcommand{\commentcolor}{\color{Blue}}
\newcommand{\typecolor}{\color{Cyan}}
\newcommand{\elcomment}[1]{\hfill{\comment{#1}}}
\newcommand{\funcname}[1]{\texttt{#1}}%{\textrm{\color{DarkBlue}{#1}}}
\newcommand{\typename}[1]{\textbf{#1}}%{\textrm{\color{Cyan}{#1}}}    

% -----------------------------

\begin{abstract}
Reconfigurable state machine replication is an important enabler of
elasticity for replicated cloud services, which must be able to
dynamically adjust their size as a function of changing load and
resource availability. We introduce a new {\em generic framework}\/ to
allow the reconfigurable state machine implementation to be derived
from a collection of arbitrary non-reconfigurable state machines. Our
reduction framework follows the {\em black box}\/ approach, and does
not make any assumptions with respect to its execution environment
apart from reliable channels.  It allows higher-level services to
leverage {\em speculative command execution}\/ to ensure uninterrupted
progress during the reconfiguration periods as well as in situations
where failures prevent the reconfiguration agreement from being
reached in a timely fashion. We apply our framework to obtain a
reconfigurable speculative state machine from the non-reconfigurable
Paxos implementation, and analyze its performance on a realistic
distributed testbed. Our results show that our framework incurs
negligible overheads in the absence of reconfiguration, and allows
steady throughput to be maintained throughout the reconfiguration
periods.
% Our results show sustained throughput of $12$K commands/sec in the
% absence of reconfigurations, which decreases only {\em marginally}\/
% during the reconfiguration times.
\end{abstract}

\vspace*{2cm}
\noindent {\bf Eligible for the best student paper award:} Dmitri Perelman
is a full-time graduate student

\vspace*{1cm}
\noindent The paper includes {\bf 12.5}\/ pages out of which
approximately {\bf 2} are occupied by the figures.

\newpage
\pagestyle{plain}
\setcounter{page}{1}

\section {Introduction}
\label{sec:intro}

% From reconfiguration to elastic systems
Replicated state machine, or RSM~\cite{lamport-state-machine}, is an
important tool for maintaining integrity of distributed applications
and services in failure-prone data center and cloud computing
environments. In these settings, the infrastructure needs to adapt to
changing resource availability, load fluctuations, variable power
consumption, and data locality constraints. In order to meet these
requirements, RSM must support {\em reconfiguration}, i.e., dynamic
changes to replica set, or quorum system. It is essential to ensure
that reconfiguration incurs minimum disruption to availability and
performance, in order to enable building truly elastic services.  The
ability to perform reconfigurations in a non-disruptive fashion
provides system designers with a powerful paradigm that can enable
many optimizations.  This includes proactive replacement of suspected
or slow nodes at low cost, adapting to the changing environment
characteristics (e.g. network delay, or diurnal load fluctuations, and
many others).

% Nontrivial ^ 2
Reconfigurable RSM has been proposed in multiple contexts (e.g.,
~\cite{ReconfStateMachine,VerticalPaxos,stoppable}).  Each solution
implemented a slightly different set of requirements, and all proved
nontrivial. Designing this functionality for a realistic environment
with minimal impact on performance is even more challenging. Na\"ive
constructions follow the ``brick-wall approach'' in which the flow of
user commands is stalled until the new configuration is installed and
the state is transferred to it.  Ideally, systems should strive to
avoid this, and favor implementations with near-seamless hand-on, that
maintain steady throughput and latency during the transition periods.

%First, the reconfiguration protocol must ensure that no new user commands are executed in the new configuration
%before it has been agreed by the members of the old one (\emph{pipeline stall}). 
%Secondly, the state transfer mechanism should guarantee that every command is successfully delivered to new configurations (\emph{state %transfer stall}).
%Finally, moving a configuration might incur costly ``warm up'' algorithms, like the first phase in Paxos, which is
%executed by a newly elected leader upon a configuration change (\emph{startup stall}). 
%The above mentioned factors might degrade system throughput and responsiveness, which essentially prevents 
%the use of highly-frequent reconfiguration policies in practice. 

% We rock -- both generic and efficient
This paper introduces a framework for constructing reconfigurable
state machines from collections of non-reconfigurable ones. We follow
the {\em black box}\/ approach that assumes nothing about the
execution environment except the existence of reliable communication
channels. Our reduction is both simple and generic, i.e., the
underlying RSM implementation is completely opaque to the framework.
Furthermore, it does not compromise efficiency, incurring negligible
overhead in the absence of reconfigurations and avoiding system stalls
upon reconfiguration.  

The main ideas underlying our framework are as follows. Each newly
proposed configuration is associated with its own instance of RSM, and
all active RSM are executed concurrently to each other.  The globally
consistent {\em trunk}\/ of commands is created by gluing together the
totally ordered command sequences produced by each RSM. When switching
from one RSM to another, the latter is chosen based on the outcome of
the configuration agreement in the former. Our framework also relieves
RSMs of state transfer responsibilities by ensuring that the latest
trunk is transferred to the new configuration concurrently with the
RSM execution. This way, each newly created RSM is completely
independent from its predecessor, and in particular, can start
executing from its initial state. We leverage this capability in our
Paxos-based reconfigurable RSM implementation to supply each newly
created Paxos instance with the identifier of a deterministically
chosen leader thus eschewing the first phase of Paxos if the
configured leader does not fail.

% Digging deeper ...
Another important optimization made possible by the RSM independence
is the ability to {\em speculatively}\/ overlap their execution with
the reconfiguration protocol, thus considerably reducing the command
latency during reconfiguration periods. Specifically, each RSM is made
available for accepting commands for the new configuration as soon as
it is proposed, and without waiting for it to be agreed upon by the
parent RSM. The proposals associated with the new configuration
proceed to be ordered concurrently with the reconfiguration agreement,
and are added to the trunk as soon as the configuration is agreed
upon. This way, our framework allows unbounded degree of parallelism
in the command execution during the transition periods avoiding the
performance problems of the ``brick-wall'' solutions. In addition, the
benefits of speculation become more substantial as the network delay
grows thus making speculative solutions attractive in wide-area
network settings.

% ... and deeper
%The technique for avoiding system stalls for state transfer is streaming the 
%commands decided upon in past configurations lazily, in parallel with voting on 
%new proposals. Thus, a new branch in our framework starts working without waiting 
%for the state transfer of its predecessor to complete. In other words, the reconfiguration 
%rate does not depend on the amount of data to be transferred upon each reconfiguration. 
%Our framework defines rigorous rules for garbage-collecting of obsolete branches, 
%in a way that allows a process to precisely identify the moment its data has been 
%transferred to other replicas, and can be safely released.

The modularity of our framework enables a range of additional features
useful in practical settings. One such feature is supporting rolling
software upgrades: i.e., the implementation of the deployed RSMs can
be replaced with the newer one without stopping the system. Likewise,
misbehaving or buggy RSMs can be restarted or replaced on-the-fly with
the minimum impact on the system operation as per the
recovery-oriented computing (ROC)~\cite{roc} guidelines.

% \frappe\/ ({\em Fast
% ReplicAtion Platform for Elastic services}).  

% In the context of Paxos algorithm we also show that the use of
% independent branches for different configurations eliminates the need
% for running the first phase of Paxos upon a new configuration startup.
% This observation is true even in the cases where the new branch does
% not share the same leader with its preceding branch.  This is due to
% the fact that the initial state of each replica in a new configuration
% $C$ is deterministically defined by $C$'s unique id, including the
% initial accepted values, which are implicitly known to the new leader.

We used our framework to implement a full-fledged reconfigurable
replication platform using non-reconfigurable
Paxos~\cite{classic-paxos,paxos-made-simple} as its underlying
non-reconfigurable RSM, and experimentally studied its performance.
The results demonstrate that our system achieves high throughput and
low latency in the absence of reconfigurations, which stay almost
unchanged under highly dynamic reconfiguration
scenarios. Specifically, the throughput is unaffected in the runs with
reconfiguration rate of $5$ per second, and degrades only by 20\% when
reconfiguration rate achieves that of $20$ per second. In addition,
our study indicates that the command latency in the vicinity of
reconfiguration stays the same as that in the absence thereof.

In summary, the contributions made by our work are as follows:
\begin{itemize}
\item new generic framework for constructing reconfigurable RSM from
non-reconfigurable ones, which is simple, modular, and efficient;
\item new speculative approach to enable overlapping command ordering
with the reconfiguration protocol thus reducing the reconfiguration
latencies;
\item practical implementation that leverages our framework to
transform Paxos-based RSM to a highly efficient reconfigurable RSM;
\item detailed experimental study of the implementation above
demonstrating the benefits of our framework in practical settings.
\end{itemize} 

The rest of the paper is organized as follows.  Related work is
discussed in Section~\ref{sec:relwork}.  System model is presented in
Section~\ref{sec:model}. Section~\ref{sec:smr} introduces
non-reconfigurable and reconfigurable variants of RSM, and rigorously
specify their properties. In Section~\ref{sec:algo}, we present our
reduction algorithm, and argue its correctness. The replication
platform built on top of our framework, and its performance are
discussed in Section~\ref{sec:perf}. Section~\ref{sec:conc} concludes
the paper.

\section {Related work}
\label{sec:relwork}

Modular approaches to specifying RSM and their constituent building
blocks have been addressed in several prior works. Boichat et
al.~\cite{DBLP:journals/sigact/BoichatDFG03} introduced modular
decomposition of Paxos into weak leader election and round-based
consensus abstraction, and \cite{DBLP:journals/dc/ChocklerM05} studied
the abstraction of ranked register capturing the essence of the Paxos
consensus algorithm. These papers however, do not address
reconfiguration. Stoppable Paxos~\cite{stoppable} presents a framework
for combining non-reconfigurable Paxos instances into a reconfigurable
RSM, but does not attempt to abstract away the internals of the
underlying Paxos implementation. In addition, \cite{stoppable} does
not support concurrent execution of the individual Paxos instances.

Several approaches to alleviating reconfiguration bottleneck in
reconfigurable state machines have been proposed. The original idea by
Lamport, described
in~\cite{classic-paxos,paxos-made-simple,ReconfStateMachine}, and
implemented in SMART~\cite{Lorch_thesmart}, was to delay the effect of
the configuration agreed in a specific consensus instance by a fixed
number $\alpha$ of successive consensus instances. If the
configuration must take effect immediately, the remaining instances
can be skipped by passing a special ``window closure'' decree
consisting of $\alpha$ consecutive {\em noop}\/ instances. Although
this approach allows up to $\alpha$ consecutive commands to be
executed concurrently, choosing the right value of $\alpha$ is
nontrivial. On the one hand, choosing $\alpha$ to be too small may
under-utilize the available resources. On the other hand, large values
of $\alpha$ may not match the actual service reconfiguration rate
resulting in too frequent invocations of the window closure decrees
(which must complete synchronously).

Chubby~\cite{Burrows:2006:CLS:1298455.1298487} and
ZooKeeper~\cite{Hunt:2010:ZWC:1855840.1855851} expose
high-level synchronization primitives (respectively, locks and
watches) that can be used to implement a reconfigurable state machine
within the client groups. The solutions based on this approach are
however, vulnerable to timing failures, and therefore, either restrict
their failure model~\cite{chain-rep}, or rely on additional
synchronization protocols within the replication groups themselves to
maintain consistency~\cite{VerticalPaxos,spinnacker}.

Vertical Paxos~\cite{VerticalPaxos} removes the configuration
agreement overhead from the critical path by delegating it to an
auxiliary ``configuration master''. The reconfiguration involves an
extra step of synchronizing with the read quorums of all preceding
configurations causing throughput degradation. In addition, the
configuration master itself is implemented using the $\alpha$-based
reconfigurable Paxos protocol, and therefore, suffers from the
limitations similar to those discussed above.

Dynamic reconfiguration has been extensively studied in the context of
virtually synchronous group
communication~\cite{aleta,birman-joseph,esti,lynch}, and
reconfigurable read/write registers~\cite{Lynch02rambo:a,shraer}. The
reconfiguration protocols described in these papers do not aim to
support consistency semantics as strong as those of state machine
replication, and therefore do not directly apply in our context.
Birman et al.~\cite{VirtuallySynchronousMethodology} present a
replication framework unifying reconfigurable state machine and
virtual synchrony in which the normal operation is suspended during
the reconfiguration periods.

Optimistic and speculative approaches to mask the coordination latency
have been extensively studied in the past in a variety of contexts
(such as e.g., group communication~\cite{bankomat}, and database
replication~\cite{tutu}).  However, to the best of our knowledge,
speculative reconfigurable state machine replication studied in this
paper has not yet been addressed in the prior work.

% \input{prelim}
%\vspace{-0.4cm}
% \input{algorithm}
%\vspace{-0.4cm}

\section{System Model}
\label{sec:model}

We consider an asynchronous message-passing system consisting of the
(possibly infinite) set of processes $P$. Each pair of processes is
connected by a point-to-point FIFO channel. For simplicity, we assume
the {\em crash}\/ failure model: i.e., the process experiencing a
crash failure stops executing any further instructions forever.  A
process that never crashes throughout the system execution is called
{\em correct}.  The channels are reliable in the sense that a message
sent by a correct process $p$ to another correct process $q$ is
eventually received by $q$. Extending our techniques to support
stronger benign failure models (such as crash/recovery, message
omission, and network partitions and merges) is straightforward, and
our prototype implementation described in Section~\ref{sec:perf} is
capable of doing that.

\section{State Machine Replication}
\label{sec:smr}

In this section, we introduce the notions of non-reconfigurable and
reconfigurable {\em Replicated State Machines (RSM)}, and specify
their properties. Our specification style loosely follows the I/O
automata formalism of~\cite{ioa}.

For the following we will fix $\Command$ to be the set of the user
commands, and $\Config$ be the set of the configuration
identifiers. Each configuration $C$ is associated with a finite set of
processes in $P$, denoted $\members(C)$.

\subsection{Non-Reconfigurable RSM (\nrrsm)}

\begin{figure}[!t]
%\floatname{algorithm}{Figure}
\label{alg:nrrsm-api}
%\centering
%\begin{multicols}{2}
\centering
\fbox{\begin{minipage}{0.9\textwidth}
\begin{distribalgo}
%\scriptsize
\INDENT{{\bf Inputs:}}
\STATE $\join_{C,p}$
\STATE $\propose(cmd)_{C,p}$, $cmd\in \Command$
\ENDINDENT

\INDENT{{\bf Outputs:}}
\STATE $\learn(cmd)_{C,p}$, $cmd\in \Command$
\ENDINDENT
\end{distribalgo}
\end{minipage}}
%\end{multicols}
\caption{The $\nrrsm(C)_p$, $C\in \Config$, signature for process
$p\in \members(C)$:}
\end{figure}

The {\em Non-Reconfigurable Replicated State Machine (\nrrsm)}\/ is
parametrized by a configuration $C\in \Config$. The interface
supported by $\nrrsm(C)$ for each process $p\in \members(C)$ appears
in Figure~\ref{alg:nrrsm-api}. $\nrrsm(C)$ accepts the commands $cmd
\in \Command$ submitted by the $C$'s members through the
$\propose(cmd)_C$, requests and outputs a totally ordered sequence of
$\learn(cmd)_C$ notifications.

The process $p$'s instance of $\nrrsm(C)$, denoted $\nrrsm(C)_p$, is
activated by the $\join_{C,p}$ request. $\nrrsm(C)$ becomes
operational once sufficiently many members $p$ of $C$ have executed
the $join_{C,p}$ request. The precise initial membership depends on
the \nrrsm\/ implementation being used, desired resiliency, failure
model, and environment properties. For example, if $\nrrsm$ is
implemented using the non-reconfigurable Paxos
algorithm~\cite{paxos-made-simple}, $\nrrsm(C)$ may become operational
as soon as a majority of $members(C)$ have executed $join_{C,p}$.  We
do not explicitly model the process ``leaves'' since they are
equivalent to the process crashes in the crash failure model.

% Note that although we do not model process leaves, this does not
% affect the generality of our specification. In any realistic RSM
% implementation, the replica that left, and later rejoined is expected
% to reconstruct the latest system state upon rejoining; and the replica
% that left and never rejoined is equivalent to a crashed
% replica. Therefore, leaves do not address any interesting, otherwise
% not captured, behavior, and can be omitted.

% In the following, we will often omit the process subscript from the
% state machine actions if it is clear from the context.

We now specify the set of properties that must be satisfied by the
correct implementations of \nrrsm. The properties are defined in terms
of the sequences of actions $\alpha$ of its external interface in
Figure~\ref{alg:nrrsm-api}. We start by specifying the environment
assumptions that must hold in order for the \nrrsm\/ implementation to
be correct:

\begin{property} {\em (Well-Formedness)}

\begin{itemize}

\item If $e=\propose_{C,p}$ is an event in $\alpha$, then there exists
$\join_{C,p}$ that precedes $e$ in $\alpha$.

\item For each process $p\in \members(C)$, at most one $\join_{C,p}$
occurs in $\alpha$.

\item For each command $cmd\in \Command$, at most one
$\propose(cmd)_{\ast,\ast}$ occurs in $\alpha$.

\end{itemize}
\label{prop:well-form-nr}
\end{property}

Next, we specify the $\nrrsm(C)$ safety properties.  The Integrity
property below asserts the following two facts: (1) a process cannot
output any \learn\/ events before it has joined $\nrrsm(C)$, and (2)
each learnt command must be previously proposed. Formally,

\begin{property}[Integrity]
Let $e=\learn(cmd)_{C,p}$ be an event in $\alpha$. Then, the following
holds:

\begin{itemize}

\item There exists $\join_{C,p}$ event that precedes $e$ in $\alpha$.

\item There exists $\propose(cmd)_{C,q}$ event that precedes $e$ in $\alpha$.

\end{itemize}
\label{prop:integr-nr}
\end{property}

The next property states that each command is learnt at most once by
each process:

\begin{property}[No Duplication]
For each command $cmd\in \Command$ and process $p$, there exists at
most one $\learn(cmd)_p$ event in $\alpha$.
\label{prop:nodup-nr}
\end{property}

The following property requires that the commands are learnt in the
same order by all members of $C$. Formally:

\begin{property}[Linearizability]
There exists a sequence of commands $\bar{x}=x_1, x_2, \dots$ such that for
each process $p\in \members(C)$, if $\bar{\pi}=\pi_1, \pi_2, \dots$ is the
sequence of the $\learn_p$ events output by $p$ in $\alpha$, then, for
all $i\ge 1$, $\pi_i=\learn(x_i)_{C,p}$.
\label{prop:linear-nr}
\end{property}

% We say that $\alpha$ satisfies {\em safety}\/ iff it satisfies both
% the NR-Integrity and NR-Linearizability properties above.

We now turn to specifying the \nrrsm\/ liveness. The property below
asserts that after some time $t$ (which is the parameter of the
property), each command proposed by a correct member of $C$ is
eventually learnt by all correct members of $C$ provided they have
joined $\nrrsm(C)$. Formally, \nrrsm\/ is said to be {\em live}\/
after time $t$ in $\alpha$ if the following holds:

\begin{property}[Liveness]
If $\propose(cmd)_p$ is an event occurring at a correct process $p$
after $t$, then for each correct process $q\in \members(C)$ such that
$\join_{C,q}$ is an event in $\alpha$, there exists
$\learn(cmd)_{C,q}$ that occurs at time $t' > t$ in $\alpha$.
\label{prop:nr-live}
\end{property}

% We are now ready to define what it means for the $\nrrsm(C)$
% implementation to be correct:

% \begin{definition}
% The automaton $A$ whose external interface consists of the actions in
% Figure~\ref{alg:nrrsm-api} is a {\em correct}\/ implementation of
% $\nrrsm(C)$ iff each well-formed execution $\alpha$ of $A$ satisfies
% Properties~\ref{prop:integr-nr}, \ref{prop:linear-nr}, and
% \ref{prop:nr-live}.
% \end{definition}

\subsection{Reconfigurable RSM (\rrsm)}

\begin{figure}[h]
%\floatname{algorithm}{Figure}
\label{alg:rrsm-api}
\centering
\fbox{\begin{minipage}{0.9\textwidth}
\begin{distribalgo}
%\scriptsize
%\begin{multicols}{2}
\INDENT{{\bf Inputs:}}
\STATE $\propose(C, cmd)_p$, $C\in \Config$, $cmd\in \Command$, $p\in \members(C)$
\STATE $\recon(C, C')_p$, $C, C' \in \Config$, $p\in \members(C)$
\ENDINDENT
\INDENT{{\bf Outputs:}}
\STATE $\learn(cmd)_p$, $cmd\in \Command$
\STATE $\newconf(C)_p$, $C\in \Config$
\STATE $\ready(C)_p$, $C\in \Config$, $p\in \members(C)$
\ENDINDENT
%\end{multicols}
\end{distribalgo}
\end{minipage}}
\caption{The $\rrsm$ Signature for Process $p\in P$:}
\end{figure}

The $\rrsm$ external interface appears in Figure~\ref{alg:rrsm-api}.
In contrast to the non-reconfigurable RSM, the configuration of {\em
Reconfigurable RSM (\rrsm)}\/ can be changed dynamically through the
user inputs. Initially, the \rrsm's configuration is the distinguished
configuration $C_0$. Afterwards, new configurations are proposed
through the $\recon(C,C')$ requests, which include two parameters: a
known configuration $C$, and the target configuration $C'$ to which
$C$ is requested to reconfigure. 

The \recon\/ requests can only be initiated at one of the members of
$C$. The members of $C'$ respond with $\ready(C')$ events once they
are aware of $C'$, and ready to accept proposals associated with
$C'$. If $C'$ is eventually accepted as the configuration to follow
$C$ in the global order, the members $q$ of both $C$ and all of the
subsequent configurations (including $C'$) will be notified of that
through the $\newconf(C')_q$ events.

Similarly to \nrrsm, \rrsm\/ accepts commands submitted through the
$\propose$ requests, and outputs $\learn$ notifications for the
commands whose slot in the total order has been finalized.  In
addition, each \propose\/ request also includes the configuration
parameter that indicates the desired configuration to be used to order
the provided command.

Note that the \rrsm\/ interface allows the proposals (of both commands
and configurations) to be issued against configurations, which have
been reported as ready, but whose ordering has not yet been completed
(and may subsequently fail). This provides implementations with
flexibility in choosing how to deal with the proposals that have been
received by \rrsm\/ while reconfiguration is still in progress.  In
our implementation (see Section~\ref{sec:algo}), these proposals are
ordered {\em speculatively}\/ using a separate instance of the
non-reconfigurable state machine being executed concurrently with the
reconfiguration protocol. As we show in Section~\ref{sec:perf}, this
leads to substantial reductions in the reconfiguration delays provided
the proposed configuration will be eventually included into the total
order (which is a common occurrence in practice). The implementation
can limit or completely eliminate speculative behavior by tweaking the
timing of the \ready\/ reports.

% conflicting proposals may prevent a configuration, which has
% been reported as ready, from being eventually finalized and included
% into the totally ordered stream of commands and configurations
% produced by \rrsm. The proposals issued against ready, but
% non-finalized configurations (to which we refer as {\em speculative})
% could however be leveraged by the implementation to mask
% reconfiguration delays as we explain in Section~\ref{sec:algo}.

\begin{figure}[!t]
        \centering
        \fbox{\begin{minipage}{0.5\textwidth}
	\centering
	\includegraphics[height=4cm]{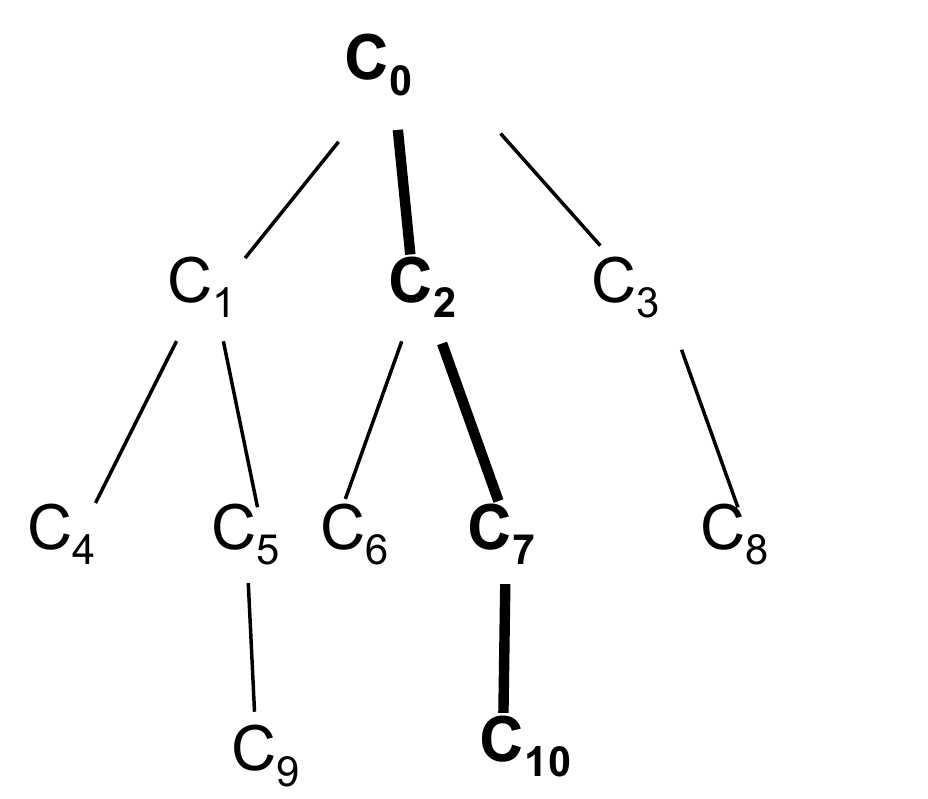}
        \end{minipage}}
	\caption{The configuration tree formed by the \recon\/
        events. The trunk configurations are highlighted.}
	\label{fig:conf-tree}
\end{figure}

All configurations submitted in the course of the \rrsm\/ execution
$\alpha$ form the {\em tree}\/ with the root $C_0$, and $C$ being the
parent of $C'$ for all $C, C' \in \Config$ iff $\recon(C,C')$ has
occurred in $\alpha$ (see Figure~\ref{fig:conf-tree}). The
configurations reported in the \newconf\/ notifications form a
continuous path, called the {\em trunk}, in the configuration tree
starting from $C_0$.

As before, we specify the correctness properties for the \rrsm\/
implementation in terms of the sequences of actions $\alpha$ of its
external interface in Figure~\ref{alg:rrsm-api}. We start by defining
the well-formedness conditions, which capture the permissible
interaction patterns between the \rrsm\/ and its environment:

\begin{property} {\em (Well-Formedness)}
\begin{itemize}

\item If $e \in \{\propose(C, \ast)_p, \recon(C, \ast)_p\}$ is an
event in $\alpha$, then there exists $\ready(C)_p$ that precedes $e$
in $\alpha$ unless $C=C_0$

\item For each command $cmd\in \Command$, there exists at most one
$\propose(cmd)_{\ast}$ in $\alpha$.

\item For each configuration $C\in \Config$, there exists at most one
$\recon(\ast,C)_{\ast}$ in $\alpha$\footnote{Note that only the
proposed configuration {\em identifiers}\/ are required to be unique,
not the set of members associated with those identifiers.}.

\end{itemize}
\label{prop:well-form-r}
\end{property}

% Note that the above does not preclude a newly proposed configuration
% $C$ from being used {\em immediately}\/ for proposing commands and
% configurations even before its corresponding $\newconf(C)$ event is
% delivered. As we discuss in Section~\ref{sec:algo}, this allows the
% \rrsm\/ implementation to overlap the execution of the reconfiguration
% protocol with the command ordering thus speeding up the command
% delivery during the reconfiguration periods.

Next, we specify the $\rrsm$ safety properties.  The Integrity
property below asserts that each \learn\/ (resp., \newconf)
notification delivered at a process $p$ must be preceded by both: (1)
$p$'s join, and (2) $\propose(C,\ast)$ ($\recon(C,\ast)$) where $C$ is
the latest configuration that has been delivered at $p$ before the
notification event. Formally:

\begin{property}[Integrity]
Let $e\in \{\learn(cmd)_p,\newconf(C')_p,\ready(C')_p\}$, be an event
in $\alpha$, and $C$ be the configuration delivered by the latest
$\newconf$ event that precedes $e$ at $p$. Then, all of the following
holds:

\begin{itemize}

\item If $e=\learn(cmd)_p$, then there exists $q\in
\members(C)$ such that $\propose(C, cmd)_q$ (resp., $\recon(C,C')_q$)
is an event in $\alpha$.

\item If $e\in \{\newconf(C')_p,\ready(C')_p\}$, then there exists
$q\in \members(C)$ such that $\recon(C,C')_q$) is an event in
$\alpha$.

\end{itemize}
\label{prop:integr-r}
\end{property}

The next property states that each command (resp., configuration) is
learnt (resp., delivered) at most once by each process:

\begin{property}[No Duplication]
For each $cmd \in \Command$, there exists at most one $\learn(cmd)_p$,
and for each $C\in \Config$, there is at most one $\ready(C)_p$, and
at most one $\newconf(C)_p$ event that occurs at each process $p\in P$
in $\alpha$.
\label{prop:nodup-r}
\end{property}

The most important safety property supported by \rrsm\/ is {\em
linearizability}, which requires that all processes deliver the same
sequence of \learn\/ and \newconf\/ events. Formally,

\begin{property}[Linearizability]
There exists a sequence of commands and configurations $\bar{x}=x_1, x_2,
\dots$ such that for each process $p\in P$, if $\bar{\pi}=\pi_1, \pi_2,
\dots$ is the sequence of the $\learn_p$ and $\newconf_p$ events
output by $p$ in $\alpha$, then for all $i\ge 1$, if $x_i\in \Config$,
then $\pi_i=\newconf(x_i)_p$, and if $x_i\in \Command$, then
$\pi_i=\learn(x_i)_p$.
\label{prop:linear-r}
\end{property}

% The Information Propagation
% property below states that each command (resp., configuration) learnt
% (resp., delivered) by a correct process is eventually learnt (resp.,
% delivered) by all correct processs that have joined \rrsm. Formally,

% \begin{property}[Information Propagation]
% Let $p$ and $q$ be correct processes such that $\learn(cmd)_p$ (resp.,
% $\newconf(C)_p$) is an event occurring at time $t$ in $\alpha$.  Then,
% there exists time $t'\ge t$ such that all correct processes $q$ that
% have previously joined \rrsm\/ deliver $\learn(cmd)_q$ (resp.,
% $\newconf(C)_q$) before $t'$ in $\alpha$.
% \label{prop:propagation}
% \end{property}

We now specify the $\rrsm$ liveness. The following property says that
if every proposed configuration includes at least one correct member,
then eventually, the system stabilizes in the sense that (1) each
command proposed by a correct process is eventually learnt provided
new reconfigurations attempts cease to be initiated, and (2) each
proposal to reconfigure a previously installed configuration
eventually triggers a \newconf\/ event. Formally,

\begin{property}[Liveness]
Suppose that every configuration proposed in $\alpha$ includes at least
one correct member. Then, there exists Global Stabilization Time (GST)
such that for all times $t>GST$, if $C_0, C_1, \dots, C_k$ is the
sequence of configurations induced by the longest sequence of
$\newconf$ events delivered by some process before $t$, then, both of
the following holds:
\begin{itemize}

\item If no $\recon(C,\ast)$ events with $C=C_k$ occur after $GST$,
then for each correct $p \in \members(C_k)$, if $\propose(C_k,cmd)_p$
is an event that occurs after $t$ in $\alpha$, then it is eventually
followed by $\learn(cmd)_q$ at all correct $q\in \members(C_k)$.

\item There exists configuration $C$ such that if a correct $p\in
\members(C_k)$ invokes $\recon(C_k, \ast)_p$ after $t$, then it is
eventually followed by $\newconf(C)_q$ at all correct $q\in
\members(C_k) \cup \members(C)$.

\end{itemize}

\label{prop:live-r}
\end{property}

In addition, we require that each $\recon(\ast, C)$ invoked by a
correct process is eventually followed by the $\ready(C)$ notification
delivered at each correct member of $C$. 

\begin{property}
If $\recon(\ast,C)_p$ is an event that occurs at a correct process $p$
in $\alpha$, then it is followed by $\ready(C)_q$ for each correct
$q\in \members(C)$.
\label{prop:ready-live}
\end{property}

\section{Reduction Algorithm}
\label{sec:algo}

In this section, we present the implementation of the algorithm to
transform a collection of $\textsc{nr-rsm}(C)$, $C\in \Config$ to
$\textsc{r-rsm}$, and argue its correctness. The implementation is the
composition of the automata $\rrsm_p$ for each process $p\in P$. The
code executed by each $\rrsm_p$ appears in Algorithm~\ref{alg:rrsmp}.

\newcounter{alg:tree-pseudo:lines}
\begin{algo*}[!ht]
\caption{$\rrsm$ from $\{\nrrsm(C)\}$. The $\rrsm_p$ code.}
\label{alg:rrsmp}

%\centering
\scriptsize
\begin{multicols}{2}
\begin{distribalgo}[1]

\INDENT {\textbf{Types}}
  \STATE $\Config$: the set of configuration identifiers with the initial identifier $C_0$
  \STATE $\Command$: the set of command indetifiers
\ENDINDENT

\medskip

\VAR 
  \STATE $\status \in \{active, idle\}$
  \STATE $\view \subseteq P$
  \STATE $\branches$: mapping from $\Config$ to $\seqOf(\Config \cup \Command)\cup \{\bot\}$ \label{line:branches}
  \STATE $\trunk$: $\seqOf(\Config \cup \Command)$ \label{line:trunk-def}
  \STATE $\next \in \Nat$
  \STATE $\curConf \in \Config$ \label{line:curconf-def}
\ENDVAR

\medskip

\INIT
  \STATE $\branches(C) \gets \bot$ for all $C\in \Config$
  \STATE $\trunk \gets []$
  \STATE $\next \gets 0$
  \STATE $\curConf \gets C_0$ \label{line:curconf-init}
  \STATE $\view \gets \emptyset$
  \IF {$p\in \members(C_0)$}
        \STATE $\joinConf(C_0)$
        \STATE $\view \gets \members(C_0)$
  \ENDIF
\ENDINIT

\medskip

% \INDENT{$\textbf{\leave}$:}
%         \STATE $status \gets idle$
%         \STATE $view \gets \emptyset$
%         \FORALL {$C$ : $\branches(C) \neq \bot$}
%                 \STATE $\branches(C) \gets \bot$
%                 \STATE $\leave_C$
%         \ENDFOR
%         \STATE Disable the State Transfer task
% \ENDINDENT

% \medskip

\INDENT{$\textbf{\propose}(C, cmd)$:}
        \IF {$\branches(C)\neq \bot \wedge \branches(C)[i]\not\in \Config$\/ 
        for all $0\le i \le length(\branches(C))$}
                \STATE $\propose(cmd)_{C}$ \label{line:prop1}
        \ENDIF
\ENDINDENT

\medskip

\INDENT{$\textbf{\recon}(C, C')$:}
        \IF {$\branches(C)\neq \bot \wedge \branches(C)[i]\not\in \Config$\/ for all 
        $0\le i \le length(\branches(C))$}
                \FORALL {$q\in \members(C')$} \label{line:new-branch-spec-start}
                        \STATE $send(\langle\textsc{join}, C, C'\rangle)_{p,q}$
                \ENDFOR \label{line:new-branch-spec-end}
                \STATE $\propose(C')_{C}$ \label{line:prop2}
                \STATE $view \gets view \cup \{C'\}$
        \ENDIF
\ENDINDENT

\medskip

\UPON {$\learn(x)_{C}$:}
        \STATE Append $x$ to $\branches(C)$ \label{line:learn-start}
        \IF {$C=\curConf \wedge \next < length(\branches(\curConf))$}
                \STATE $\learnNext(\branches(\curConf)[\next])$
        \ENDIF \label{line:learn-finish}
\ENDUPON

\medskip

\UPON {receiving $\tup{\textsc{state}, tr, v}$ message from $q\in P$}
       \FORALL {$i=length(trunk)$ \textbf{to} $length(tr)-1$} \label{line:trunk-patch-start}
                \STATE $\learnNext(tr[i])$
       \ENDFOR \label{line:trunk-patch-end}
       \STATE $view \gets view \cup v \cup \{q\}$ \label{line:view-merge}
\ENDUPON

\medskip

\UPON {receiving $\tup{\textsc{join}, C, C'}$ message from $q\in \members(C)$}
        \IF {$\branches(C')=\bot$} \label{line:new-branch-start}
                \STATE $\joinConf(C')$
                \STATE Output $\ready(C')$
                \STATE $\view \gets \view \cup \members(C) \cup \members(C')$
        \ENDIF \label{line:new-branch-finish}
\ENDUPON

% \INDENT {\hl{\hbox{\textbf{upon} receiving $\langle\textsc{new-branch}, C\rangle$ message from $q\in P$}}}
%         \INDENT {\hl{\hbox{\textbf{if} $status=joined \wedge \branches(C)=\bot$}} \label{line:new-branch-start}}
%                 \STATE \hl{$\branches(C) \gets []$}
%                 \STATE \hl{$\join_{C}$}
%                 \STATE \hl{$\report(C)$}
%         \ENDINDENT 
%         \STATE \hl{\hbox{\textbf{end if}}} \label{line:new-branch-finish}
% \ENDINDENT
% \STATE \hl{\hbox{\textbf{end upon}}}

\medskip

\TASK {State Transfer}  \label{line:state-transfer-start}
        \IF {$\status=active$}
        \INDENT {Periodically:}
                \FORALL {$q\in view$}
                        \STATE $send(\tup{\textsc{state}, trunk, view})_{p,q}$ \label{line:send-state}
                \ENDFOR
        \ENDINDENT
        \ENDIF
\ENDTASK \label{line:state-transfer-end}

\medskip

\PROCEDURE {$\learnNext(x)$, $x\in \Command \cup \Config$} \label{line:learn-next-start}
        \STATE Append $x$ to trunk \label{line:append-trunk}
        \IF {$x\in \Config$} 
                \IF {$\branches(x) = \bot \wedge p\in \members(x)$} \label{line:new-sm}
                        \STATE $\joinConf(x)$
                        \STATE Output $\ready(C)$
                        \STATE $\view \gets \view \cup \members(x)$ \label{line:view-learn-next}
                \ENDIF \label{line:trunk-conf-end}
                \STATE Output $\newconf(x)$ \label{line:trunk-conf-start}
                \STATE $\next \gets 0$
                \STATE $\curConf \gets x$
        \ELSE
                \STATE Output $\learn(x)$ \label{line:learn-cmd-start}
                \STATE $\next \gets \next + 1$ \label{line:learn-cmd-finish}
        \ENDIF
\ENDPROC \label{line:learn-next-end}

\medskip

\PROCEDURE {$\joinConf(C)$, $C\in \Config$}
         \IF {$\status=idle$}
                \STATE $\status \gets active$
         \ENDIF
         \STATE $\branches(C) \gets []$
         \STATE $\join_{C}$  \label{line:join}
\ENDPROC

\setcounter{alg:tree-pseudo:lines}{\value{ALC@line}} % store the line number
\end{distribalgo}
\end{multicols}
\end{algo*}

% \begin{observation}
% Each command or configuration $x$ is associated with a unique
% configuration $C$ where $C$ is the first argument of the $\propose$ or
% $\recon$ requests through which $x$ was submitted to $\rrsm$.
% \label{observ:1}
% \end{observation}

% \begin{observation}
% All configurations submitted in the course of a well-formed execution
% $\alpha$ of \rrsm\/ form the {\em tree}\/ having the initial
% configuration $C_0$ as its root, and $C$ as the parent of $C'$ for all
% $C, C' \in \Config$ iff $\recon(C,C')$ is an event in $\alpha$.
% \label{observ:2}
% \end{observation}

The implementation of $\rrsm$ can be viewed as consisting of the
following three phases: 

\begin{description}

\item [Configuration-specific ordering:] Each proposed configuration
$C$ is associated with the non-reconfigurable state machine
$\nrrsm(C)$, which is used to order the commands and configurations
associated with $C$. The local outputs of each $\nrrsm(C)_p$ are
stored in the $\branches_p$ data structure
(line~\ref{alg:rrsmp}.\ref{line:branches}), which maps each $C$ to the
sequence of commands and configurations output by $\learn_{C,p}$
events.

\item [Configuration tree pruning:] The branches of the configuration
tree in Figure~\ref{fig:conf-tree} are {\em pruned}\/ so that the
surviving configurations form a single consistent {\em trunk}. To
ensure that all processes prune the branches in a consistent fashion,
the successor $C'$ of each configuration $C$ is determined by the
output of $\nrrsm(C)$.

\item [Total order construction:] Each process $p$ concatenates the
totally ordered fragments stored in $\branches_p$ in the order induced
by the configuration trunk to produce a globally-consistent sequence
of commands and configurations, which is stored in the $\trunk_p$
variable. $\trunk_p$ is then iterated to produce a globally consistent
sequence of the \learn\/ and \newconf\/ outputs.

\end{description}

The above three phases are executed concurrently, with the
configuration-specific ordering phase being executed concurrently for
each individual configuration. Below we describe the implementation of
each of the above phases in more detail.

\paragraph{Configuration-Specific Ordering:} A process $p$ joins
$\nrrsm(C)$ when it either receives the $\tup{\textsc{join},C}$
message
(lines~\ref{alg:rrsmp}.\ref{line:new-branch-start}--\ref{alg:rrsmp}.\ref{line:new-branch-finish}),
or encounters $C$ in the course of the total order construction phase
(lines~\ref{alg:rrsmp}.\ref{line:new-sm}--\ref{alg:rrsmp}.\ref{line:trunk-conf-end}).
From that point on, all commands and configurations $x$ submitted by
process $p$ through either $\propose(C,x)_p$ or $\recon(C,x)_p$ are
forwarded to the $p$'s local instance of $\nrrsm(C)$ through the
$propose(x)_{C,p}$ requests (see lines
\ref{alg:rrsmp}.\ref{line:prop1} and
\ref{alg:rrsmp}.\ref{line:prop1}). Each ordered command or
configuration delivered by the $\learn_{C,p}$ event is then appended
to $\branches(C)_p$.

The state machines created upon the reception of the
$\textsc{join}$ messages are used to {\em speculatively order}\/
the proposals corresponding to the configurations, which have not yet
been incorporated into the global total order, thus reducing the
reconfiguration time. We discuss speculation in more detail in
Section~\ref{sec:spec} below.

\paragraph{Configuration Tree Pruning:}
The configuration tree is pruned by selecting the {\em first}\/
configuration appearing in $\branches(C)_p$ to be the $C$'s successor
by all processes $p\in \members(C)$. Since all entries of
$\branches(C)_p$ appear in the same order at all processes $p\in
\members(C)$, they all will choose the same configuration $C's$ as the
$C$'s successor. All other processes will learn of $C'$ through state
transfer (see below). The successor chosen for each configuration $C$
becomes explicit when it is incorporated into $\trunk_p$ in the
course of the total order construction phase (lines
\ref{alg:rrsmp}.\ref{line:trunk-conf-start}--\ref{alg:rrsmp}.\ref{line:trunk-conf-end}).

\paragraph{Total Order Construction:}
The chain of the surviving branches currently known to each process
$p$ is kept in the $trunk_p$ variable (see
line~\ref{alg:rrsmp}.\ref{line:trunk-def}). In addition, $\curConf_p$
(see line~\ref{alg:rrsmp}.\ref{line:curconf-def}) holds the
configuration corresponding to the last branch on
$trunk_p$. Initially, $\curConf_p=C_0$
(line~\ref{alg:rrsmp}.\ref{line:curconf-init}) so that $trunk_p$
consists of a single branch corresponding to the root of the
configuration tree $C_0$.

As the entries are added to $\branches(\curConf_p)$
(lines~\ref{alg:rrsmp}.\ref{line:learn-start}--\ref{alg:rrsmp}.\ref{line:learn-finish}),
they are copied to $trunk_p$
(line~\ref{alg:rrsmp}.\ref{line:append-trunk}) and learnt one-by-one
(line~\ref{alg:rrsmp}.\ref{line:learn-cmd-start}--\ref{alg:rrsmp}.\ref{line:learn-cmd-finish})
until a new configuration $C$ is encountered. At this point,
$\curConf_p$ is reassigned to $C$ thus choosing
$\branches(\curConf_p)$ as the next branch to feed $trunk_p$ (lines
~\ref{alg:rrsmp}.\ref{line:trunk-conf-start}--\ref{alg:rrsmp}.\ref{line:trunk-conf-end}).

Since as we argued above, $C$ is the only configuration that can be
chosen as the $\curConf_p$'s successor at all processes $q\in P$,
$\branches(C)$ can be the only one selected to follow
$\branches(\curConf_q)$ in $\trunk_q$. We therefore, proved the
following:

\begin{lemma}
All well-formed executions of the $\rrsm$ implementation in
Algorithm~\ref{alg:rrsmp} satisfy Property~\ref{prop:linear-r}
\label{lem:linear-r}
\end{lemma}

\subsection{State Propagation}

Since for each configuration $C$, the commands and configurations
ordered by $\nrrsm(C)$ are only delivered to the members of $C$, we
need an additional mechanism to ensure they propagate down the
configuration trunk. This mechanism is implemented by the {\em State
Transfer}\/ task
(lines~\ref{alg:rrsmp}.\ref{line:state-transfer-start}--
\ref{alg:rrsmp}.\ref{line:state-transfer-end}), which is executed in
the background at each process $p$ that have joined at least one of
the proposed configurations. 

It proceeds by continuously gossipping its entire $\trunk_p$
(line~\ref{alg:rrsmp}.\ref{line:send-state}) to the set of processes
being accumulated in the $\view_p$ variable\footnote{Note that in
reality the state transfer can be made considerably more efficient
through a variety of techniques, such as e.g., push-pull
gossip\cite{scuttlebutt}, and log contraction~\cite{ReconfStateMachine}}.
By line \ref{alg:rrsmp}.\ref{line:view-learn-next}, $\view_p$ is
guaranteed to include the members of all configurations comprising the
configuration trunk known to $p$. In particular, this includes the
members of the configuration $C'$ that immediately follows $C$ in
$trunk_p$. If $p$ is correct, all correct members of $C'$ will
eventually receive $\trunk_p$ that will include the entire sequence of
commands ordered within $C$ up to and inclusively of $C'$. Moreover,
once there is a correct $q\in \members(C')$ that delivers
$\newconf(C')$, all correct members of both $C'$ and its immediate
successor (if any) are guaranteed to receive the entire trunk known to
the members of $C$ even if all members of $C$ stop gossipping their
trunks.  We therefore, have the following:

\begin{lemma}[State Propagation]
Let $C$ and $C'$ be two configurations such that $C'$ follows $C$ in
the configuration trunk. Then, if there exists a correct member $p$ of
$C$ that continues to gossip $\trunk_p$ until $\newconf(C')$ is
delivered by a correct member $q$ of $C'$, then all correct members of
$C'$ will eventually incorporate all commands and configurations
ordered within $C$ into their trunks.
\label{lem:propagation}
\end{lemma}

The result above establishes criteria that can be used in practice for
determining when the members of old configurations can be taken off
line without compromising the system liveness. For example, suppose
that a majority of the members of each installed configuration do not
crash, unless they are taken off line in an orderly fashion. Then, for
each configuration $C$, a minority of $\members(C)$ can be taken off
line once a majority of $\members(C)$ deliver $\newconf(C')$
notification for some configuration $C'$; and the remaining members
can be disconnected once a majority of $C'$ delivers $\newconf(C')$.

\subsection{Liveness}

The liveness of our implementation follows from the liveness of its
constituent non-reconfigurable state machines and the State
Propagation property above. 

Specifically, assume that all state machines instantiated in the
course of the execution of the code in Algorithm~\ref{alg:rrsmp}
become live (in the sense of Property~\ref{prop:nr-live}) after some
time $t$. Suppose that $C$ is the last configuration of the
system. Since all configurations include at least one correct process,
by Lemma~\ref{lem:propagation}, there is a time $t'>t$ such that all
correct members of $C$ will incorporate $C$ into their trunks and join
$\nrrsm(C)$ before $t'$.

If no new reconfiguration attempts are made from this point onwards,
every command $cmd$ proposed by a correct member of $C$ through
$\propose(C,cmd)$ is guaranteed to appear in $\branches(C)_q$,
followed by $\trunk_q$, until it eventually output through
$\learn(cmd)_q$. Otherwise, there will be $C'$ such that $\recon(C,
C')$ will result in $C'$ to be included into $\trunk_p$ of all correct
processes in $C$, from where by Lemma~\ref{lem:propagation}, it is
guaranteed to eventually propagate to all correct members of
$C'$. Hence, we proved the following:

\begin{lemma}
Each well-formed execution $\alpha$ of the \rrsm\/ implementation in
Algorithm~\ref{alg:rrsmp} satisfies Property~\ref{prop:live-r}
provided all $\nrrsm$ implementations instantiated in $\alpha$ are
live after some $t$.
\end{lemma}

\subsection{Speculation}
\label{sec:spec}

One important property of our \rrsm\/ implementation is that the
non-reconfigurable state machine for each configuration $C$ is
completely {\em independent}\/ from the state machines for other
configurations, and could proceed ordering the commands and
configurations submitted under $C$ {\em concurrently}\/ with them. In
particular, since the members of $C$ are notified of the new
configuration immediately upon receiving its corresponding
reconfiguration request
(line~\ref{alg:rrsmp}.\ref{line:new-branch-spec-start}--
\ref{alg:rrsmp}.\ref{line:new-branch-spec-end}), this also applies to
the commands submitted under $C$ while the successor of its parent
configuration is still being chosen.  These commands will be available
for incorporating into the trunk and subsequent learning as soon as
$C$ becomes the chosen successor of its parent configuration, thus
reducing (or eliminating altogether) reconfiguration delays.

The precise degree of the performance improvement depends on the
choice of the underlying non-reconfigurable state machine
implementation, and the network parameters (primarily, the
communication delay). As we show in Section~\ref{sec:perf}, for the
Paxos-based $\rrsm$ implementation, speculation helps to eliminate
penalties associated with throughput and latency during
reconfiguration times, making them almost indistinguishable from those
achieved during the normal operation.
% Morever, the relative throughput
% improvement grows with the network delay thus making the speculative
% solutions particularly appealling in wide-area network settings.

%\input{arch}

%\input{paxos}

\section{Implementation and Evaluation}
\label{sec:perf}

We applied our reduction framework to implement a full-fledged
reconfigurable replication platform using non-reconfigurable
Paxos~\cite{classic-paxos,paxos-made-simple} as its underlying \nrrsm\/
implementation.

To shield the users from intricacies of selecting viable
configurations, our replication platform augments \rrsm\/ with two
additional modules: Configuration Manager (CM) and Command Queue
(CQ). CM implements the logic to detect failures, and produce new
configurations, which are then submitted to \rrsm\/ through the
\recon\/ requests. The configuration membership is based on the
failure suspicions and user policies. An example policy may ask for
every configuration to include sufficiently many healthy nodes.  CQ is
responsible for associating user commands with configurations and
submitting them to the platform through the \propose\/ requests.  The
configurations are chosen based on the \ready\/ notifications thus
leveraging the speculation capabilities of \rrsm\/ to reduce
reconfiguration delays. 

Each Paxos instance created by \rrsm\/ is supplied with the identifier
of a deterministically chosen leader thus avoiding the overheads of
the first phase of Paxos if the configured leader does not fail. Thus,
normally, only the second phase of Paxos is needed to complete each
individual agreement instance resulting in a single round-trip delay
to order a command.

We studied the performance of our implementation experimentally using
the testbed comprised of 4 IBM HS22 blades equipped with Intel Xeon
X5670 processor with 24 2.93GHz cores and 64GB RAM. Each machine was
equipped with 1GB network card, and ran Red Hat Linux. A single
replica was hosted on each machine.

The replicas were subjected to request streams generated in either
{\em synchronous}\/ or {\em asynchronous}\/ fashion. In the
synchronous mode, the requests were issued from multiple threads each
of which was waiting for the response to the previously submitted
request to arrive before submitting the new one. The synchronous mode
allowed us to exercise high degree of control over the offered system
load by varying the number of simultaneously executed synchronous
threads. In contrast, in the asynchronous mode, each thread was
submitting requests as they were generated without waiting for the
prior requests completion. The asynchronous mode was used to drive the
system to its maximum utilization. 

\begin{figure}[!t]
        \centering
%        \begin{minipage}{0.5\textwidth}
	\centering
	\includegraphics[height=6cm]{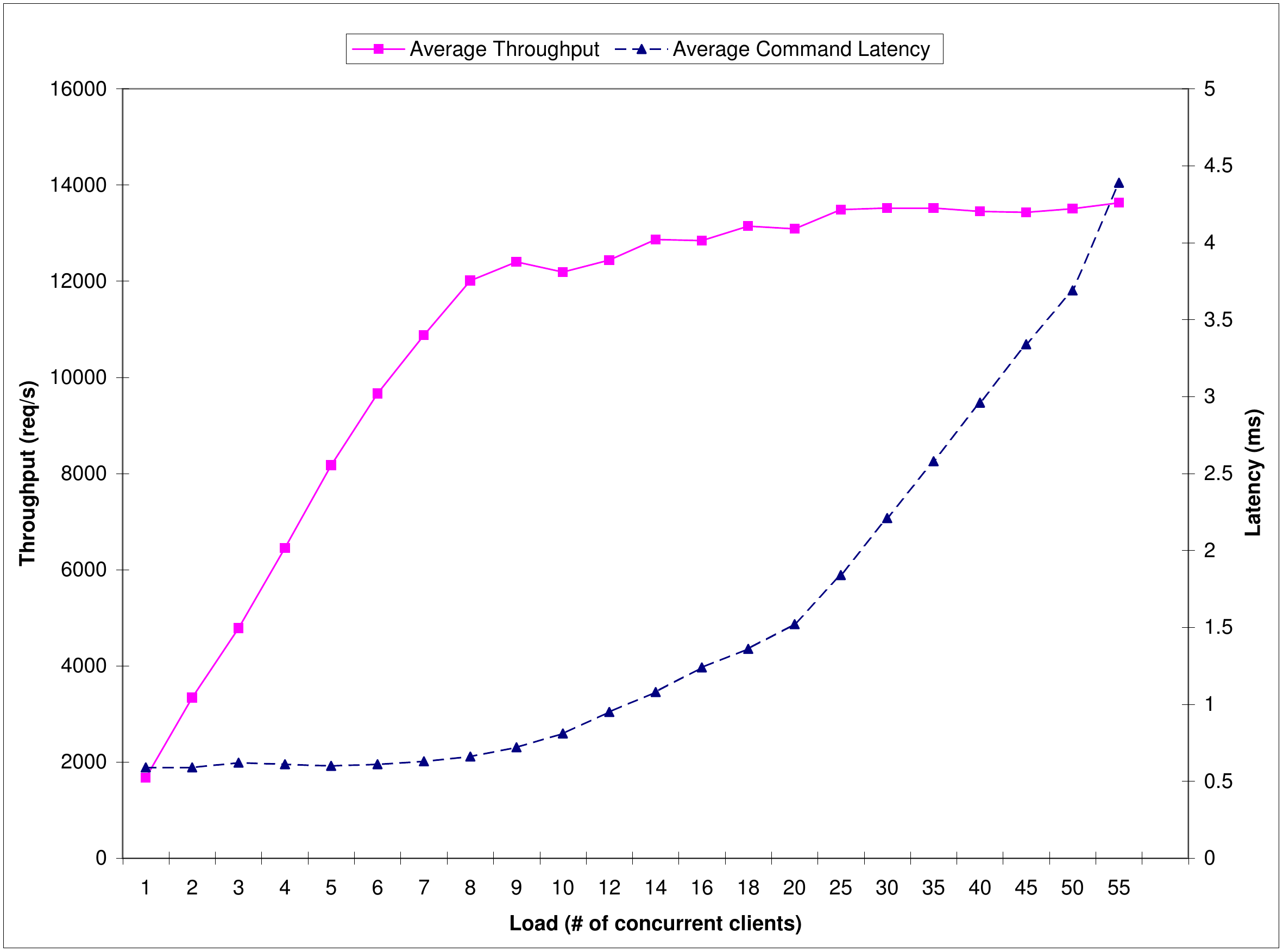}
%        \end{minipage}
	\caption{Throughput and Latency in the Absence of Reconfiguration.}
	\label{fig:normal}
\end{figure}

In the first experiment, we studied the throughput and latency as a
function of the offered system load in the absence of reconfiguration
using a single configuration consisting of 3 replicas. The results
(see Figure~\ref{fig:normal}) show that our system is capable of
achieving the sustained throughput of 12K request/second before the
latency starts to rapidly grow. 

\begin{figure}[!t]
        \centering
%        \begin{minipage}{0.5\textwidth}
	\centering
	\includegraphics[height=6cm]{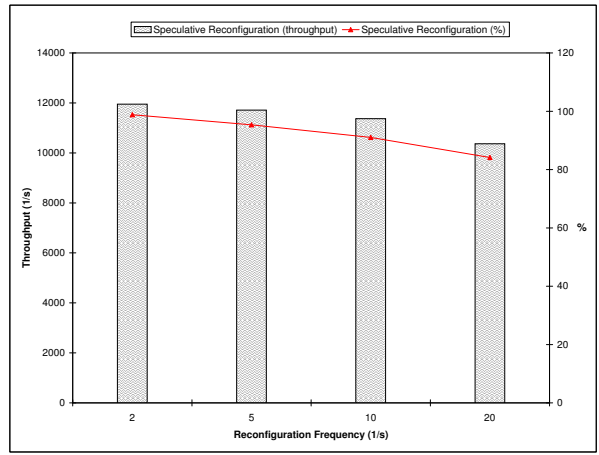}
%        \end{minipage}
	\caption{Throughput With Varied Reconfiguration Frequency.}
	\label{fig:D}
\end{figure}

In the next experiment, we studied the impact of our speculative
reconfiguration mechanism on the command throughput. For that, the
system was subjected to a series of reconfiguration requests submitted
at varied frequency. The normal commands were submitted using 10
synchronous threads, which corresponds to the maximum sustained system
load of 12K requests/second (see Figure~\ref{fig:normal}).  The
measurements are depicted (see Figure~\ref{fig:D}) as absolute command
throughput in the presence of reconfiguration, and percentage of
degradation relative to the maximum sustained throughput (12K).  The
results show that the throughput in the presence of reconfigurations
is almost identical to that achieved in the absence thereof, reaching
maximum 20\% of degradation when the system is reconfigured 20
times/second.

\begin{figure}[!t]
        \centering
%        \begin{minipage}{0.5\textwidth}
	\centering
	\includegraphics[height=6cm]{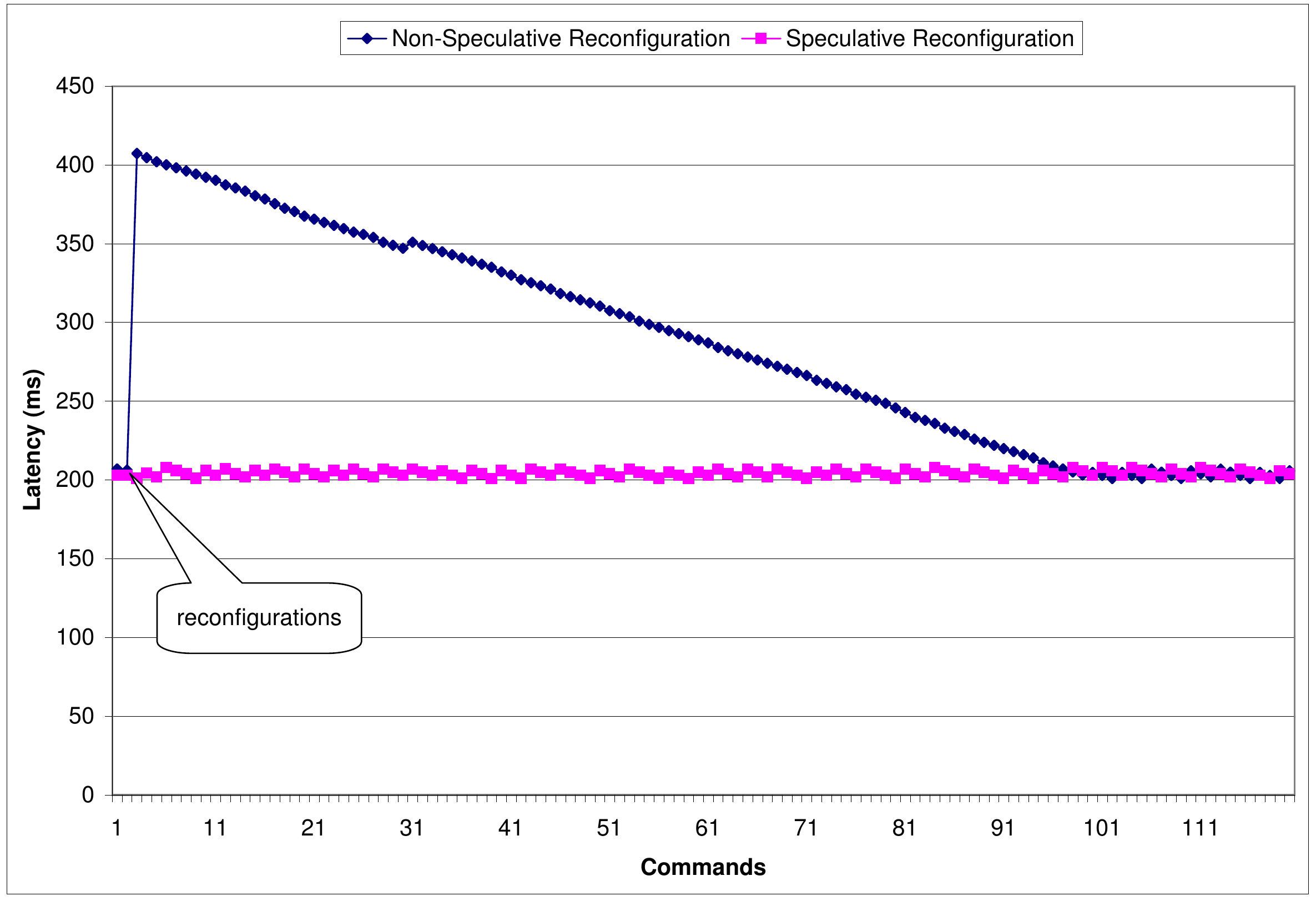}
%        \end{minipage}
	\caption{Command Latency in the Vicinity of
        Reconfiguration. The network delay was simulated to be 100ms
        on average. }
	\label{fig:latency}
\end{figure}

In the last experiment, we studied the degree of improvement produced
by speculation in terms of the latencies of the commands submitted in
close proximity of the reconfiguration requests. The results (see
Figure~\ref{fig:latency}) indicate that with speculation, the latency
is unaffected by reconfiguration staying closely to that of the normal
mode (i.e., in the absence of reconfiguration) throughout the entire
reconfiguration period. In contrast, without speculation, the latency
increases sharply for the first command, and then keeps decreasing
slowly until reaching the normal mode value in the proximity of the
100th command.

\section{Conclusions}
\label{sec:conc}

We have introduced a modular framework for transforming a collection
of arbitrary non-reconfigurable replicated state machines (RSM) into
the reconfigurable RSM implementation. Our framework follows the {\em
black box}\/ approach, and does not make any assumptions with respect
to its execution environment apart from reliable channels. The
individual state machines instantiated by our implementation can be
executed concurrently to one another, in particular, overlap each
other execution in a {\em speculative}\/ fashion to mask the
reconfiguration delays. We have applied our framework to build a
prototype of an end-to-end dynamic replication platform using
non-reconfigurable Paxos as its underlying building block, and studied
its performance on a realistic distributed testbed. Our results
demonstrated that our our platform achieves high throughput and low
latency in the absence of reconfigurations, which stay almost
unchanged under highly dynamic reconfiguration scenarios.

Our platform is being developed at IBM Research, and slated to be
included into the future IBM's platform-as-a-service offerings as a
foundational tool. In the future, we intend to explore the ways to
extend our speculation-based approach to support optimistic
replication in wide-area network settings, and augment it with an
on-line optimization framework to facilitate selection of
configurations based on the changing network parameters.

%\vspace{-0.4cm}
%\input{discussion}
%\vspace{-0.4cm}
%\input{Acknowledgments}
% \vspace{-0.2cm}
% This work has been partially supported by the EU IST project CoMiFin
% FP7-ICT-225407/2008.

\bibliographystyle{abbrv}

%\balance
\newpage
\bibliography{references}

%\appendix

%\input{appb}

\end{document}